\documentclass[amsmath,amssymb,showkeys,superscriptaddress,floatfix]{revtex4}
\usepackage{graphicx,epstopdf}
\usepackage{subfig}
\usepackage{xcolor}
\usepackage[colorlinks=true, urlcolor=blue, linkcolor=blue, citecolor=green]{hyperref}

\def\xslash{x\!\!\!\slash }

\def\vel{\left|}
\def\ver{\right|}

\begin{document}

\title{The electromagnetic multipole moments of the charged open-flavor $Z_{\bar cq}$ states}
\author{K.~Azizi}%
\email[]{kazizi@dogus.edu.tr}%
\affiliation{Department of Physics, Dogus University, Acibadem-Kadikoy, 34722 
Istanbul, Turkey}%
\affiliation{School of Physics, Institute for Research in Fundamental Sciences (IPM),
P.~O.~Box 19395-5531, Tehran, Iran}%
\author{U.~\"{O}zdem}%
\email[]{uozdem@dogus.edu.tr}%
\affiliation{Department of Physics, Dogus University, Acibadem-Kadikoy, 34722 Istanbul, Turkey}%

\date{\today}
 
\begin{abstract}
The electromagnetic multipole moments of the open-flavor $Z_{\bar cq}$ states 
are investigated by assuming 
a diquark-antidiquark picture for their 
internal structure and quantum numbers $J^{PC} = 1^{+-}$
 for their spin-parity.
In particular, their magnetic and quadrupole moments are extracted in the 
framework of light-cone QCD sum rule
by the help of the photon distribution amplitudes.
The electromagnetic multipole
moments of the open-flavor $Z_{\bar cq}$ states are 
 important 
dynamical observables, which
encode valuable information on their underlying structure.
The results obtained for the magnetic moments 
 of different structures
are considerably large 
and can be measured in future experiments.
We obtain very small values for the quadrupole moments of $Z_{\bar cq}$ states 
indicating a nonspherical charge distribution.
\end{abstract}

\maketitle

\section{Introduction}
Since 2003, there are many non-conventional
hadrons discovered experimentally, such as many XYZ
tetraquarks, $P_c^+(4380)$ and $P_c^+(4450)$ 
pentaquarks etc., which could 
not be described as the conventional hadrons composed
of two or three valence quark/antiquarks.  
They are called exotic hadrons. For some reviews on the 
 theoretical and experimental progress
 on the properties of these new states see Refs.~\cite{Nielsen:2009uh,Swanson:2006st,Voloshin:2007dx,Klempt:2007cp,Godfrey:2008nc,
Faccini:2012pj,Esposito:2014rxa,Chen:2016qju,Ali:2017jda,Esposito:2016noz,Olsen:2017bmm}.
The greatest achievement with regard to the 
exotic states was the discovery of the charged multiquark states.
The charged states with a hidden pair of heavy quark and antiquark such as 
the $Z_c^{\pm}(3900)$~\cite{Ablikim:2013mio}, $Z_c^{\pm}(4020)$~\cite{Ablikim:2013wzq}, 
$Z_c^{\pm}(4430)$~\cite{Choi:2007wga},
$Z_b^{\pm}(10610,10650)$~\cite{Belle:2011aa}, 
would be undoubtedly considered as the exotic resonances,
because these charged states cannot be explained as excited charmonium-like or bottomonium-like
states.

Most of the discovered exotic states
up to now share a common properties: 
they contain a hidden heavy quark-antiquark pair, $\bar cc$ or $\bar bb$.
However existence of the multiquarks, which do not contain
 $\bar cc$ or $\bar bb$ pairs is also possible, because fundamental
laws of QCD do not prohibit existence of such open-flavor multiquark states.
It should be noted  that they
have not been discovered experimentally, and to our best
knowledge, there are not any  candidates to be considered
for these states. They may be seen in
the exclusive reactions as the open-charm and open-bottom resonances.
In 2003, the two narrow charm-strange mesons
$D^∗_{s0}(2317)$ and $D_{s1}(2460)$ were observed in
the $D^+_s \pi^{0}$ and $D^{*+}_s \pi^{0}$
invariant mass distributions by the BABAR~\cite{Aubert:2003fg} 
and CLEO~\cite{Besson:2003cp}] collaborations, are now being
considered as candidates to open-charm tetraquark states.
In 2016, the D0 Collaboration  reported 
the observation of  a state with four different 
quark flavors, the $X(5568)$, 
and assigned the quantum numbers $J^{P}=0^+$ for it, 
but they did not 
 exclude the possibility of $J^{P} = 1^{+}$~\cite{D0:2016mwd}. 
Reported in the $B^0_s \pi^{\pm}$
final states, the $X(5568)$ meson, if exist,
cannot be categorized into the conventional meson family,
and is a good candidate of exotic tetraquark state with valence
quarks of four different flavors such as $su\bar d\bar b$ or $s d\bar u \bar b$.
The observation of these states have immediately inspired extensive
discussions on the possibility of their internal structure.
For more information see for instance Refs.~\cite{Agaev:2016mjb,Agaev:2016srl,Chen:2016spr} and references therein.
In 2017, the D0 Collaboration repeated their analysis when the $B_s$ is reconstructed semileptonically.
They reported evidence for a narrow structure,
which was consistent with their previous measurement in the hadronic decay mode~\cite{Abazov:2017poh}.
However, other experimental groups, namely the
LHCb~\cite{Aaij:2016iev}, CDF~\cite{Aaltonen:2017voc}, CMS~\cite{Sirunyan:2017ofq} and ATLAS~\cite{Aaboud:2018hgx}
collaborations could not find this resonance from analysis of their experimental data.

In order to understand the inner structure of the hadrons in the nonperturbative regime
of QCD, the main challenges are the determination of
the dynamical and statical features of hadrons 
such as their decay form factors, masses, 
electromagnetic multipole moments and so on, both experimentally and theoretically. 
Many theoretical models accurately estimate the mass and decay 
width of the discovered exotic states, but the inner structure of these states is still unclear.
In other words, the mass and decay width alone can not distinguish the inner structure of the exotic states.
Recall that the electromagnetic multipole moments are  equally important dynamical observables of the exotic states. 
The electromagnetic multipole moments include 
the spatial distributions
of the charge and magnetization in the hadrons 
and these parameters are directly related to the spatial
distributions of quarks and gluons inside the hadrons.
There are many studies in the literature devoted to the 
investigation of the electromagnetic
multipole moments of the standard hadrons,
but unfortunately relatively little are known the
 electromagnetic multipole moments of the exotic hadrons.
There are a few studies in the literature
where the magnetic dipole and quadrupole 
 moments of the exotic states are studied: see~ 
 \cite{Agamaliev:2016wtt,Ozdem:2017jqh,Ozdem:2017exj}
 for tetraquarks and
 ~\cite{Kim:2003ay,Huang:2003bu,Liu:2003ab,Wang:2005gv,Wang:2005jea,Wang:2016dzu} 
 for pentaquarks.
 More detailed analyses are needed in order to get useful 
 knowledge on the charge distribution, electromagnetic 
 multipole moments and geometric shapes of the 
 non-conventional hadrons. 
 In this study, we are going to concentrate on 
 the charged open-flavor 
 $[qq][\overline{qc}]$ states (hereafter we will denote these states as $Z_{\bar cq}$)
 with spin-parity 
$J^{PC} =1^{+-}$, and calculate their electromagnetic multipole moments 
in the framework of light-cone QCD sum rule (LCSR).
 In LCSR,  the hadronic parameters are expressed 
 in terms of the vacuum condensates and the light
cone distribution amplitudes (DAs) of the on-shell particles
(for more about this method see, 
e.g.,~\cite{Chernyak:1990ag, Braun:1988qv, Balitsky:1989ry} and references therein).

The rest of the paper is organized as follows: In section II, the calculation of the sum rules in LCSR will
be presented. In the last section, we numerically analyze the 
sum rules obtained for the multipole moments and discuss
the obtained results.
The explicit expressions of the magnetic and quadrupole moments are moved to the  Appendix A.

\section{Formalism}
In this section we derive the LCSR for the magnetic and quadrupole moments of the 
 $Z_{\bar cq}$ states. For this aim, we consider a correlation function 
in the presence of the external 
electromagnetic field ($\gamma$),

\begin{equation}
 \label{edmn01}
\Pi _{\mu \nu }(q)=i\int d^{4}xe^{ip\cdot x}\langle 0|\mathcal{T}\{J_{\mu}^{Z_{\bar cq}}(x)
J_{\nu }^{Z_{\bar cq}\dagger }(0)\}|0\rangle_{\gamma}, 
\end{equation}%
where $J_{\mu}$ is the interpolating current of $Z_{\bar cq}$ state with quantum numbers $J^{PC}=1^{+-}$
in the diquark-antidiquark picture. 
It is given in terms of
three light quark and one heavy 
quark fields as~\cite{Chen:2017rhl}:
\begin{eqnarray}
J_{\mu }^{Z_{\bar cq}}(x) &=& \Big[q_{1_a}^T(x)C\gamma_\mu q_{2_b}(x)\Big]\Big[\bar q_{3_a}(x)\gamma_5C\bar c_b^T(x)
-\bar q_{3_b}(x)\gamma_5C\bar c_a^T(x)\Big],
\label{eq:Curr}
\end{eqnarray}%
where $q_1$ is u, d and/or s-quark, $q_2$ and $q_3$ are u and/or d-quark,
$C$ is the charge conjugation matrix; and $a$ and $b$ are color indices.

In order to acquire sum rules for the magnetic and quadrupole moments, 
we need to represent the correlation function in
two different forms: (1) in terms of the quark-gluon 
parameters and distribution
amplitudes (DAs) of the photon in the deep Euclidean 
region, 
 so called the QCD representation, 
 and (2) in terms of hadronic properties,
 so called the hadronic representation.

We start our analysis by calculating the correlation 
function on  Eq.~(\ref{edmn01}) in terms of
quarks and gluon
properties in deep Euclidean region. For this purpose, 
the interpolating current is inserted 
into the correlation function and after the contracting of light and heavy quark pairs using the Wick
theorem the following result is obtained:
\begin{eqnarray}
\label{edmn11}
&&\Pi _{\mu \nu }^{\mathrm{QCD}}(q)=i
\int d^{4}xe^{ipx} \langle 0 | \Bigg\{  
\mathrm{Tr}\Big[\gamma _{5}\widetilde{S}_{c}^{b^{\prime }b}(-x)\gamma _{5}S_{q_3}^{a^{\prime }a}(-x)\Big]
\mathrm{Tr}\Big[\gamma _{\nu }\widetilde{S}_{q_1}^{aa^{\prime }}(x)\gamma _{\mu}S_{q_2}^{bb^{\prime }}(x)\Big] \notag \\
&&-\mathrm{Tr}\Big[\gamma _{5}\widetilde{S}_{c}^{a^{\prime }b}(-x)\gamma _{5}S_{q_3}^{b^{\prime }a}(-x)\Big]
\mathrm{Tr}\Big[\gamma _{\nu }\widetilde{S}_{q_1}^{aa^{\prime }}(x)\gamma _{\mu}S_{q_2}^{bb^{\prime }}(x)\Big] \nonumber\\
&&-\mathrm{Tr}\Big[\gamma _{5}\widetilde{S}_{c}^{b^{\prime }a}(-x)\gamma _{5}S_{q_3}^{a^{\prime }b}(-x)\Big]
\mathrm{Tr}\Big[\gamma _{\nu }\widetilde{S}_{q_1}^{aa^{\prime }}(x)\gamma _{\mu}S_{q_2}^{bb^{\prime }}(x)\Big] \notag \\
&&+\mathrm{Tr}\Big[\gamma _{5}\widetilde{S}_{c}^{a^{\prime }a}(-x)\gamma _{5}S_{q_3}^{b^{\prime }b}(-x)\Big]
\mathrm{Tr}\Big[\gamma _{\nu }\widetilde{S}_{q_1}^{aa^{\prime }}(x)\gamma _{\mu}S_{q_2}^{bb^{\prime }}(x)\Big]
 \Bigg\}| 0 \rangle_\gamma,
\end{eqnarray}%
where%
\begin{equation*}
\widetilde{S}(x)=CS^{\mathrm{T}}(x)C,
\end{equation*}%
with $S_{q,c}(x)$ being the quark propagators. The light and heavy propagators are given as~\cite{Balitsky:1987bk}
\begin{eqnarray}
\label{edmn12}
S_{q}(x)=S^{free}
- \frac{\langle \bar qq \rangle}{12} \Big(1-i\frac{m_{q} \xslash}{4}   \Big)
- \frac{\langle \bar qq \rangle}{192}m_0^2 x^2  \Big(1-i\frac{m_{q} \xslash}{6}   \Big)
-\frac {i g_s }{32 \pi^2 x^2} ~G^{\mu \nu} (x) \Bigg[\rlap/{x} 
\sigma_{\mu \nu} +  \sigma_{\mu \nu} \rlap/{x}
 \Bigg],
\end{eqnarray}%
and
%
\begin{eqnarray}
\label{edmn13}
S_{c}(x)=S^{free}
-\frac{g_{s}m_{c}}{16\pi ^{2}} \int_{0}^{1}dv~G^{\mu \nu }(vx)\Bigg[ (\sigma _{\mu \nu }{\xslash}
  +{\xslash}\sigma _{\mu \nu })\frac{K_{1}( m_{c}\sqrt{-x^{2}}) }{\sqrt{-x^{2}}}
+2\sigma ^{\mu \nu }K_{0}( m_{c}\sqrt{-x^{2}})\Bigg],
\end{eqnarray}%
where 
\begin{eqnarray}
&&S_{q}^{free}(x)=i \frac{{\xslash}}{2\pi ^{2}x^{4}} -\frac{m_{q}}{4 \pi^2 x^2},\nonumber\\
\nonumber\\
 &&S_{c}^{free}(x)=\frac{m_{c}^{2}}{4 \pi^{2}} \Bigg[ \frac{K_{1}(m_{c}\sqrt{-x^{2}}) }{\sqrt{-x^{2}}}
+i\frac{{\xslash}~K_{2}( m_{c}\sqrt{-x^{2}})}
{(\sqrt{-x^{2}})^{2}}\Bigg].
\end{eqnarray}
Here $K_{1,2}$ are Bessel functions of the second kind.

The correlation function contains  different types of contributions.
In first part, one of the free quark propagators in Eq.~(\ref{edmn11}) is replaced by
\begin{align}
\label{perttrans}
S^{free} (x) \rightarrow \int d^4y\, S^{free} (x-y)\,\rlap/{\!A}(y)\, S^{free} (y)\,,
\end{align}
 and the remaining three propagators are taken as the full quark propagators.

In the second case one of the light quark propagators in Eq.~(\ref{edmn11}) is replaced by
\begin{align}
\label{edmn14}
S_{\alpha\beta}^{ab} \rightarrow -\frac{1}{4} (\bar{q}^a \Gamma_i q^b)(\Gamma_i)_{\alpha\beta},
\end{align}
and the remaining propagators are taken as the 
full quark propagators, as well 
including the perturbative and the nonperturbative
contributions.
 Once Eq. (\ref{edmn14}) is plugged into Eq. (\ref{edmn11}), there appear matrix
elements such as $\langle \gamma(q)\vel \bar{q}(x) \Gamma_i q(0) \ver 0\rangle$
and $\langle \gamma(q)\vel \bar{q}(x) \Gamma_i G_{\alpha\beta}q(0) \ver 0\rangle$,
representing the nonperturbative contributions. 
The reader can find some details about the
transformations of Eqs.~(\ref{perttrans}) and (\ref{edmn14}) in Ref.~\cite{Ozdem:2017exj}.
These matrix elements 
can be written in terms of the photon DAs with definite
twists, whose expressions all can be found in Ref.~\cite{Ball:2002ps}.
The QCD side of the correlation function can be acquired in terms of QCD parameters 
 using the Eqs.~(\ref{edmn11})-(\ref{edmn14}) and after applying the Fourier transformation to 
transfer the calculations to the momentum space.

The next step is to calculate the correlation function in terms of the hadronic parameters.
To this end  we insert intermediate states of $Z_{\bar cq}$  
 with the same quantum numbers as the interpolating current into
Eq. (\ref{edmn01}). As a result, in zero-width approximation, we get

\begin{align}
\label{edmn04}
\Pi_{\mu\nu}^{Had} (p,q) = {\frac{\langle 0 \mid J_\mu^{Z_{\bar cq}} \mid
Z_{\bar cq}(p) \rangle}{[p^2 - m_{Z_{\bar cq}}^2]}} \langle Z_{\bar cq}(p) \mid Z_{\bar cq}(p+q) \rangle_\gamma
\frac{\langle Z_{\bar cq}(p+q) \mid {J^\dagger}_\nu^{Z_{\bar cq}} \mid 0 \rangle}{[(p+q)^2 - m_{Z_{\bar cq}}^2]} + \cdots,
\end{align}
where dots stand for the contributions of the higher and
continuum states and $q$ is the momentum of the photon. The matrix element
$\langle 0 \mid J_\mu^{Z_{\bar cq}} \mid Z_{\bar cq} \rangle$ is determined as
\begin{align}
\label{edmn05}
\langle 0 \mid J_\mu^{Z_{\bar cq}} \mid Z_{\bar cq} \rangle = \lambda_{Z_{\bar cq}} \varepsilon_\mu^\theta\,,
\end{align}
with $\lambda_{Z_{\bar cq}}$ being the residue of the $Z_{\bar cq}$ state.

The matrix element
$\langle Z_{\bar cq}(p,\varepsilon^\theta) \mid  Z_{\bar cq} (p+q,\varepsilon^{\delta})\rangle_\gamma$
can be written in terms of three Lorentz invariant form factors as follows~\cite{Brodsky:1992px}:
\begin{align}
\label{edmn06}
\langle Z_{\bar cq}(p,\varepsilon^\theta) \mid  Z_{\bar cq} (p+q,\varepsilon^{\delta})\rangle_\gamma
 &= - \varepsilon^\tau (\varepsilon^{\theta})^\alpha
(\varepsilon^{\delta})^\beta
\Bigg[ G_1(Q^2)~ (2p+q)_\tau ~g_{\alpha\beta}  +
G_2(Q^2)~ ( g_{\tau\beta}~ q_\alpha -  g_{\tau\alpha}~ q_\beta) \nonumber\\
&- \frac{1}{2 m_{Z_{\bar cq}}^2} G_3(Q^2)~ (2p+q)_\tau ~q_\alpha q_\beta  \Bigg]\,,
\end{align}
where $\varepsilon^{\theta}$ and  $\varepsilon^\delta$ are the 
polarization vectors of the initial and final $Z_{\bar cq}$
states and $\varepsilon^\tau$ is the polarization vector of the photon. 
The Lorentz invariant form factors $G_1(Q^2)$, $G_2(Q^2)$  and $G_3(Q^2)$ 
are related to the charge, magnetic and quadrupole form
factors through the relations
\begin{align}
\label{edmn07}
&F_C(Q^2) = G_1(Q^2) + \frac{2}{3} \eta F_{\cal D}(Q^2)\,,\nonumber \\
&F_M(Q^2) = G_2(Q^2)\,,\nonumber \\
&F_{\cal D}(Q^2) = G_1(Q^2)-G_2(Q^2)+(1+\eta) G_3(Q^2)\,,
\end{align}
where $\eta=Q^2/4 m_{Z_{\bar cq}}^2$ with $Q^2=-q^2$. 
At $Q^2 = 0 $, these form factors
are corresponding to the electric charge, 
magnetic moment $\mu$ and the quadrupole moment ${\cal D}$  
as:
\begin{align}
\label{edmn08}
&e F_C(0) = e \,, \nonumber\\
&e F_M(0) = 2 m_{Z_{\bar cq}} \mu \,, \nonumber\\
&e F_{\cal D}(0) = m_{Z_{\bar cq}}^2 {\cal D}\,.
\end{align}
Using Eqs. (\ref{edmn05})-(\ref{edmn08}) 
and imposing the condition $q\!\cdot\!\varepsilon = 0$  the Eq. (\ref{edmn04}) takes the
form,
\begin{align}
\label{edmn09}
 \Pi_{\mu\nu}^{Had} &=  \frac{ \lambda_{Z_{\bar cq}}^2}{ [m_{Z_{\bar cq}}^2 - (p+q)^2][m_{Z_{\bar cq}}^2 - p^2]}
 \Bigg[2 (p.\varepsilon) F_C(0) \Bigg(g_{\mu\nu} -\frac{p_\mu
q_\nu-p_\nu q_\mu}{ m_{Z_{\bar cq}}^2 } \Bigg) \nonumber \\
&+ F_M (0) \Bigg(q_\mu  \varepsilon_\nu - q_\nu \varepsilon_\mu +
\frac{1}{m_{Z_{\bar cq}}^2} (p.\varepsilon) (p_\mu q_\nu - p_\nu q_\mu ) \Bigg)
- \Bigg(F_C(0) + F_{\cal D}(0)\Bigg) {\frac{p.\varepsilon}{m_{Z_{\bar cq}}^2} } q_\mu
q_\nu \Bigg]\,,
\end{align}
where we inserted
\begin{align}
\sum_\lambda\varepsilon_\mu(p,\lambda) \varepsilon_\nu(p,\lambda)
=-g_{\mu\nu}+ \frac{p_\mu p_\nu}{m_{Z_{\bar cq}}^2}.
\end{align}



Equating the QCD and hadronic sides of the correlation function, we obtain
the expression of the electromagnetic multipole moments in LCSR
 in terms of the QCD degrees of freedom and 
 the photon DAs.
 We perform the double
Borel transforms with respect to the variables $p^2$ and $(p+q)^2$ on both sides of
the correlation function in order to suppress 
the contributions of the higher states and continuum, 
and use the quark-hadron duality assumption.
By matching the coefficients of the structures 
$q_\mu \varepsilon_\nu$ and  
$(\varepsilon.p) q_\mu q_\nu$, respectively 
for the magnetic and quadrupole moments, we get 
\begin{align}
 &\mu =\frac{e^{m_{Z_{\bar cq}}^2/M^2}}{\lambda_{Z_{\bar cq}}^2} \Pi_1^{QCD},\nonumber\\
 &\mathcal{ D} = m^2_{Z_{\bar cq}}\frac{e^{m_{Z_{\bar cq}}^2/M^2}}{\lambda_{Z_{\bar cq}}^2} \Pi_2^{QCD},
\end{align}
where explicit expressions of the $\Pi_1^{QCD}$ and $\Pi_2^{QCD}$ are given in Appendix A.

\section{Numerical analysis}

In this section, we numerically analyze the results of calculations for magnetic and quadrupole moments.
We use  $m_u = m_d = 0$, $m_{s}(2~GeV)=0.096^{+0.008}_{-0.004}~GeV$,  
$\overline{m}_c(m_c) = (1.28\pm0.03)\,GeV$~\cite{Patrignani:2016xqp}, 
$\langle \bar uu\rangle(1~GeV)=\langle \bar dd\rangle(1~GeV)=(-0.24\pm0.01)^3~GeV^3$ \cite{Ioffe:2005ym},
$m_0^{2}=0.8\pm0.1~GeV^2$, $\langle g_s^2G^2\rangle=0.88~GeV^4$~\cite{Nielsen:2009uh}, 
$\chi(1~GeV)=-2.85\pm0.5~GeV^{-2}$~\cite{Rohrwild:2007yt},
$\lambda_{Z_{sq\bar qc}}=7.3 \pm 1.7 \times 10^{-3}~GeV^5$, 
$\lambda_{Z_{qq\bar qc}}=7.6 \pm 1.8 \times 10^{-3}~GeV^5$, 
$m_{Z_{sq\bar qc}}=2.826^{+0.134}_{-0.157}~GeV$ and $m_{Z_{qq\bar qc}}=2.843^{+0.115}_{-0.139}~GeV$~\cite{Agaev:2017oay}.
The parameters used in the photon DAs are given in Ref.~\cite{Ball:2002ps}.

The predictions for the electromagnetic multipole moments depend on two
 auxiliary parameters; the Borel mass parameter $M^2$ and continuum threshold $s_0$. 
 Complying with the standard procedure of the 
 sum rule method the predictions on 
 the electromagnetic multipole moments should not depend on $M^2$ and $s_0$, 
 but in real computations one can only decrease their effect to a minimum.
The working interval for the continuum threshold is chosen such that the maximum
pole contribution is acquired and the results relatively weakly depend on its choices. 
Our numerical computations lead
to the interval [10-12]~$GeV^2$ for this parameter. 
The working region for $M^2$ is determined requiring that the contributions 
of the higher states and continuum are effectively suppressed.
There are two criteria for choosing working region for the Borel parameter $M^2$: 
Convergence of the operator product expansion (OPE) and pole dominance.
The requirement of the
OPE convergence results in a lower bound,
while the constraint of 
the maximum pole contribution leads to an upper bound on it.  
Our numerical calculation shows that these requirements
are satisfied in the region $3~GeV^2 \leq M^2 \leq 4~GeV^2$ and, 
the magnetic and quadrupole moments in this
region is practically independent of $M^2$.
 In Figs. 1-2, we plot the dependencies of the magnetic and quadrupole moments on $M^2$ 
 at several fixed values of the continuum threshold $s_0$.
 As is seen, the variation of the results with 
 respect to the continuum threshold 
 causes a change on the results on the magnetic and quadrupole moments of about 15\% 
  and there is a very less dependence of the quantities under consideration on the Borel parameter
in its working interval. 
Hence, one can say that the results of the magnetic and quadrupole moments are 
almost insensitive to $s_0$ and $M^2$.
%
%

%
 
 Our final results for the magnetic and quadrupole moments are given in Table I.
The errors in the results come from the variations in the  calculations of the working regions of
 $M^2$ and $s_0$ as well as the uncertainties 
in the values of the input parameters and the photon DAs. 
We also would like to note that in Table I and Figs. 1-2, the absolute values
are given since it is not possible to determine the sign of the residue from
the mass sum rules. Therefore, it is not possible to estimate the signs of the magnetic and
quadrupole moments.

 \begin{table}[t]
	\addtolength{\tabcolsep}{11pt}
	\begin{center}
      \begin{tabular}{cccc}
            \hline\hline
            ${Z_{\bar cq}}$          & $|\mu_{Z_{\bar cq}}|[\mu_N]$  & $|\mathcal{D}_{Z_{\bar cq}}|[fm^2]$   \\
            \hline\hline
	    $[sd][\overline{uc}]$          & 1.12$\pm$ 0.18           &0.0086 $\pm$ 0.0015 \\\hline
	    $[sd][\overline{dc}]$          & 0.90$\pm$ 0.13           &0.0085 $\pm$ 0.0015\\\hline
	    $[su][\overline{uc}]$          & 0.51$\pm$ 0.24           &0.0070 $\pm$ 0.0013 \\\hline
	    $[dd][\overline{uc}]$          & 1.09$\pm$ 0.17           &0.0082 $\pm$ 0.0014 \\\hline
	    $[du][\overline{uc}]$          & 0.84$\pm$ 0.31           &0.0067 $\pm$ 0.0012 \\\hline
	    $[dd][\overline{dc}]$          & 0.93$\pm$ 0.13           &0.0082 $\pm$ 0.0014 \\\hline
	    $[uu][\overline{dc}]$          & 2.05$\pm$ 0.30           &0.016$\pm$0.003 \\
	    \hline\hline
\end{tabular}
\end{center}
\caption{Results of the magnetic and quadrupole moments of $Z_{\bar cq}$ states.}
	\label{table}
\end{table}

In summary, the electromagnetic multipole moments 
of the open-flavor $Z_{\bar cq}$ states 
have been investigated by assuming that these states 
are represented as
diquark-antidiquark structure with quantum numbers 
$J^{PC} = 1^{+-}$. 
Their magnetic and quadrupole moments have been extracted in the 
framework of light-cone QCD sum rule. The electromagnetic multipole moments of 
the open-flavor $Z_{\bar cq}$ states are 
 important dynamical observables, which would 
encode important information of their underlying structure,
charge distribution and geometric shape. 
The results obtained for the magnetic moments are considerably large 
and can be measured in future experiments.
We obtain very small values for the quadrupole moments of $Z_{\bar cq}$ states 
indicating a nonspherical charge distribution. 
 %
It is easy to see that $[sd][\overline {uc}]$ and $[dd][\overline {uc}]$ states 
belong to a class of doubly charged tetraquarks that the measurements 
of their electromagnetic parameters, like  those of the $ \Delta^{++}$ baryon,  are relatively easy compared to other exotic states. 
These kind of exotic states have not been observed so far. 
We hope our predictions on the electromagnetic moments of these states together with the results of other theoretical studies on the 
spectroscopic parameters of these states will  be useful for their searches in 
future experiments  and will hep us determine exact internal structures of these non-conventional states. 

\begin{figure}
\centering
 \subfloat[]{\label{fig:1MMMsq.eps}\includegraphics[width=0.35\textwidth]{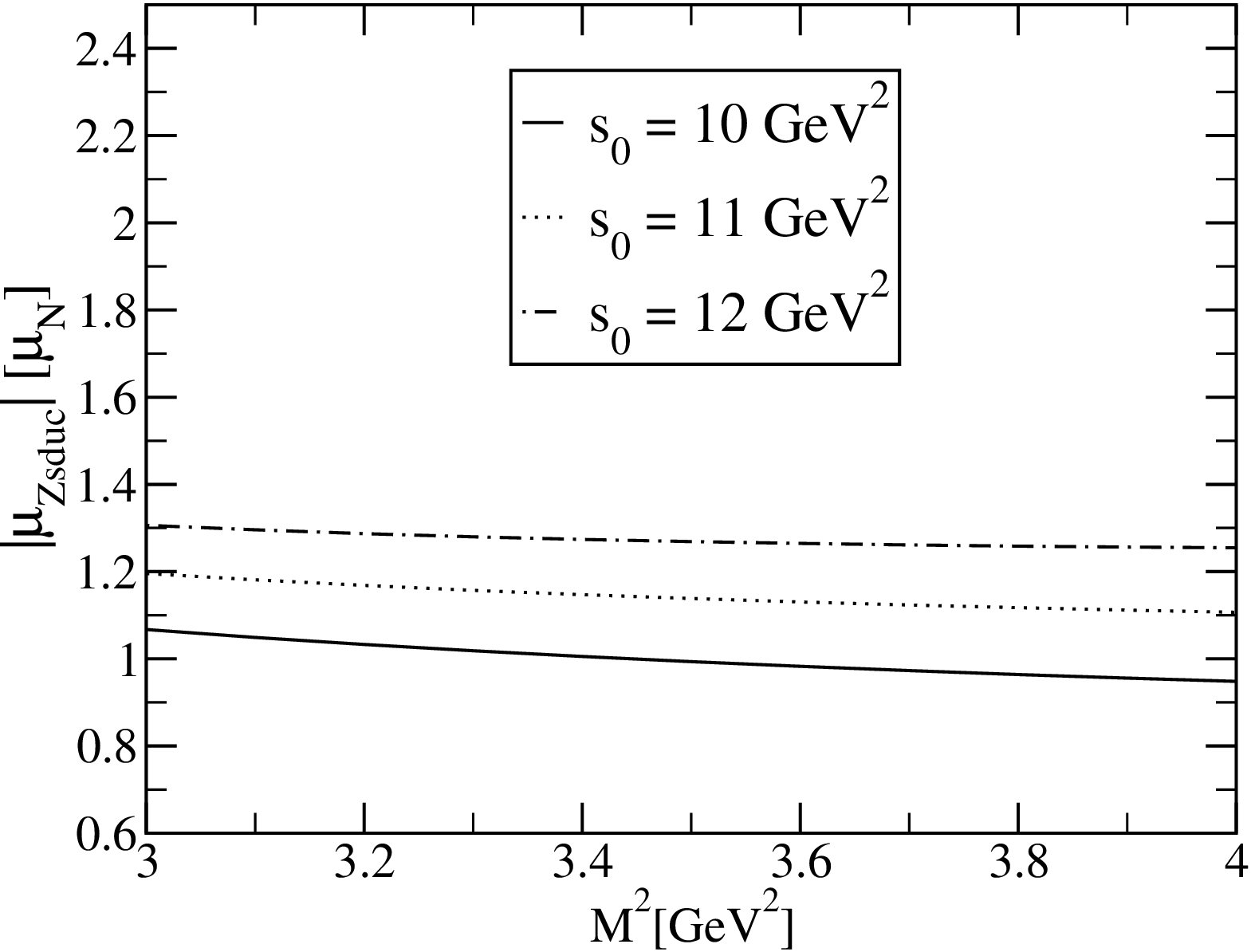}}~~~~~~~
 \subfloat[]{\label{fig:1QMMsq.eps}\includegraphics[width=0.35\textwidth]{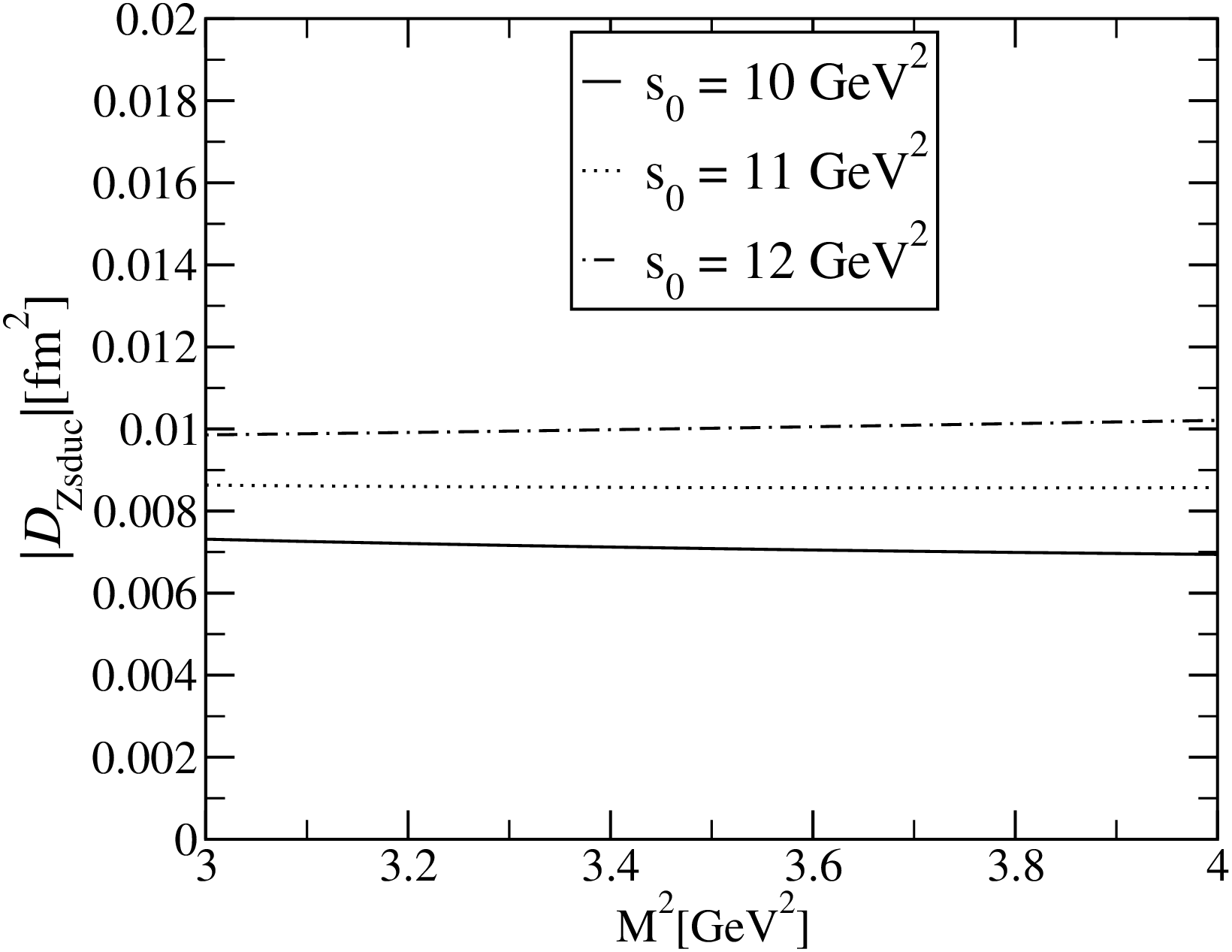}}\\
 \subfloat[]{\label{fig:2MMMsq.eps}\includegraphics[width=0.35\textwidth]{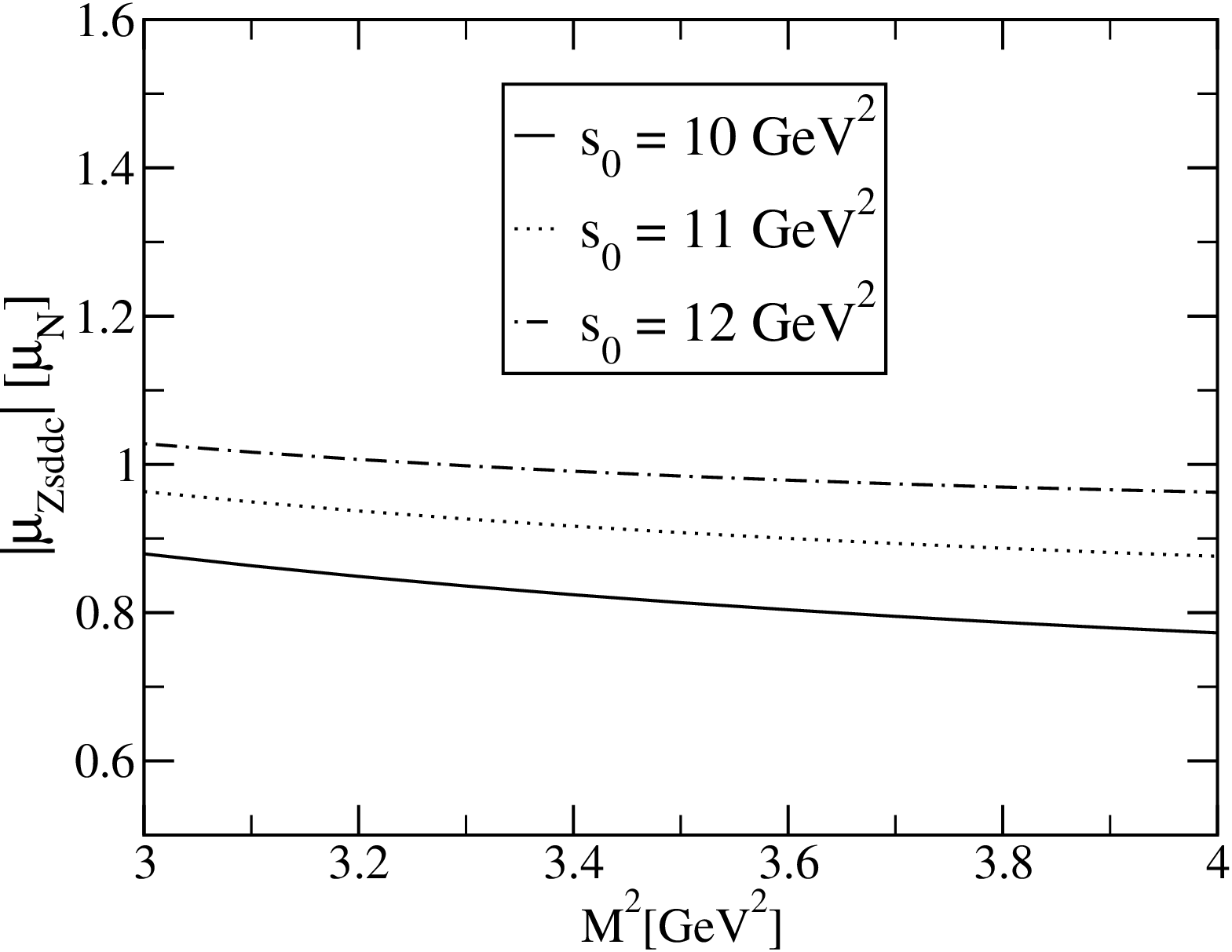}}~~~~~~~
 \subfloat[]{\label{fig:2QMMsq.eps}\includegraphics[width=0.35\textwidth]{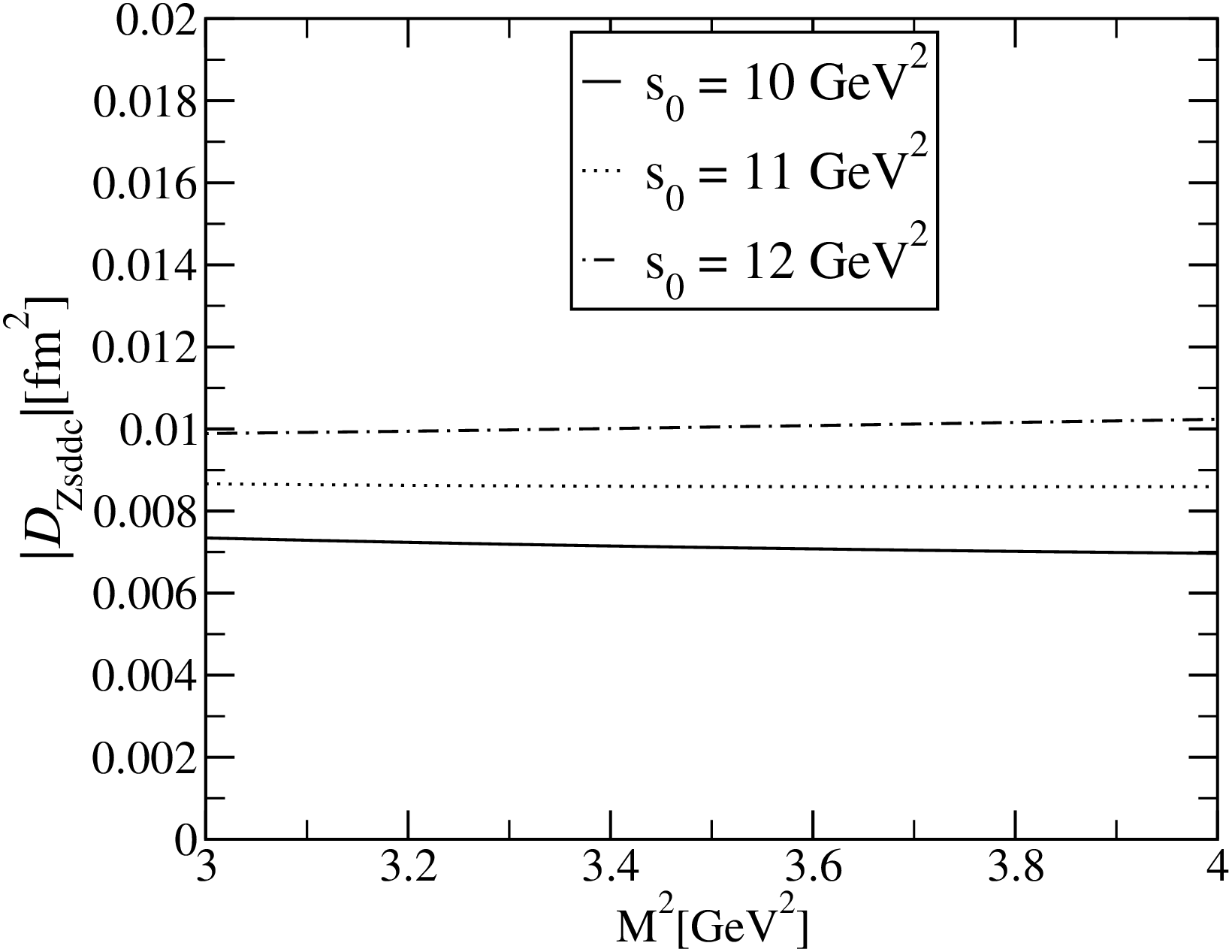}}\\
 \subfloat[]{\label{fig:3MMMsq.eps}\includegraphics[width=0.35\textwidth]{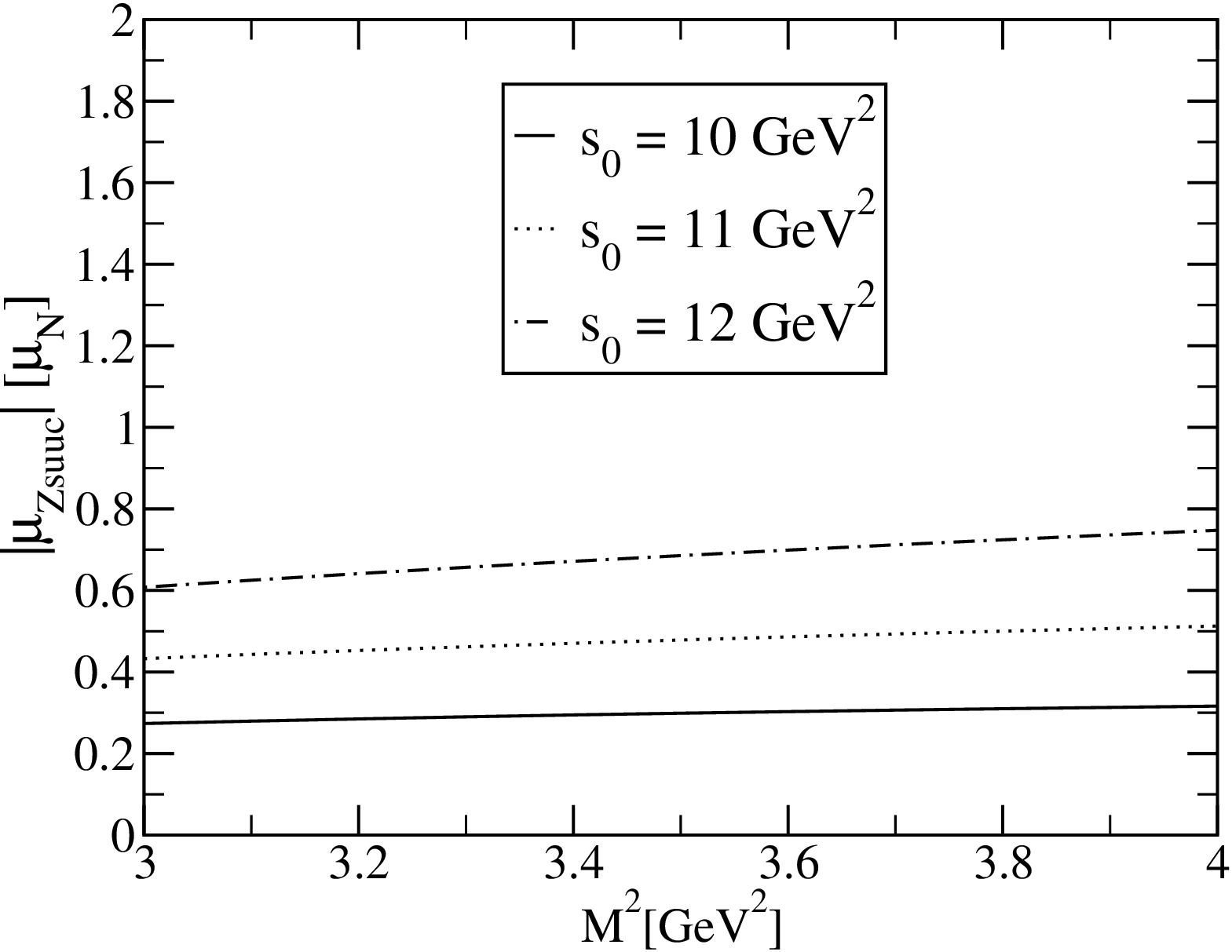}}~~~~~~~
 \subfloat[]{\label{fig:3QMMsq.eps}\includegraphics[width=0.35\textwidth]{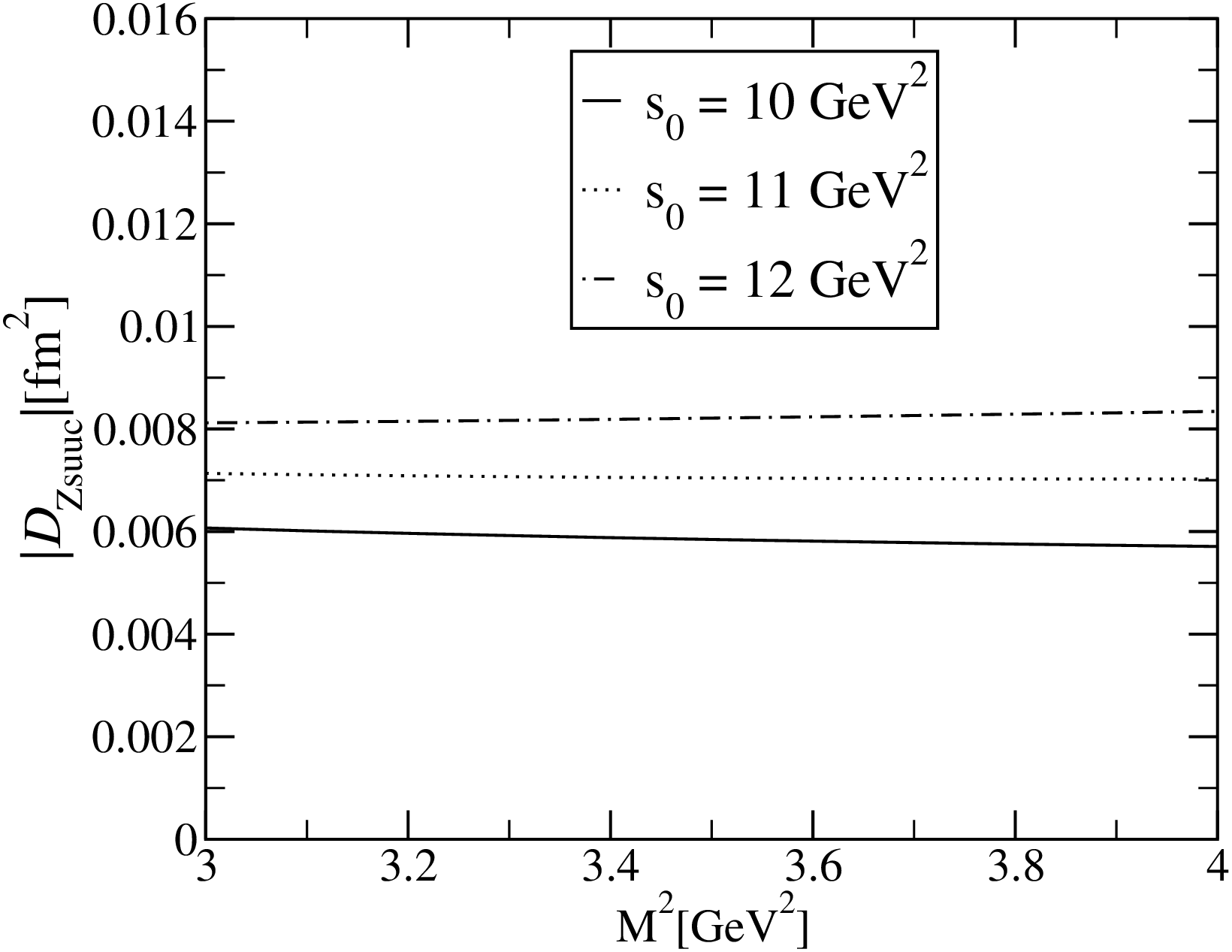}}\\
 \caption{ The dependence of the magnetic and quadrupole moments on the Borel parameter squared $M^{2}$
 at different fixed values of the continuum threshold: 
 (a) and (b) for the $Z_{sd \bar u \bar c}$ state, 
 (c) and (d) for the $Z_{sd \bar d \bar c}$ state and,
 (e) and (f) for the $Z_{su \bar u \bar c}$ state.}
  \end{figure}

  \begin{figure}
\centering
 \subfloat[]{\label{fig:4MMMsq.eps}\includegraphics[width=0.35\textwidth]{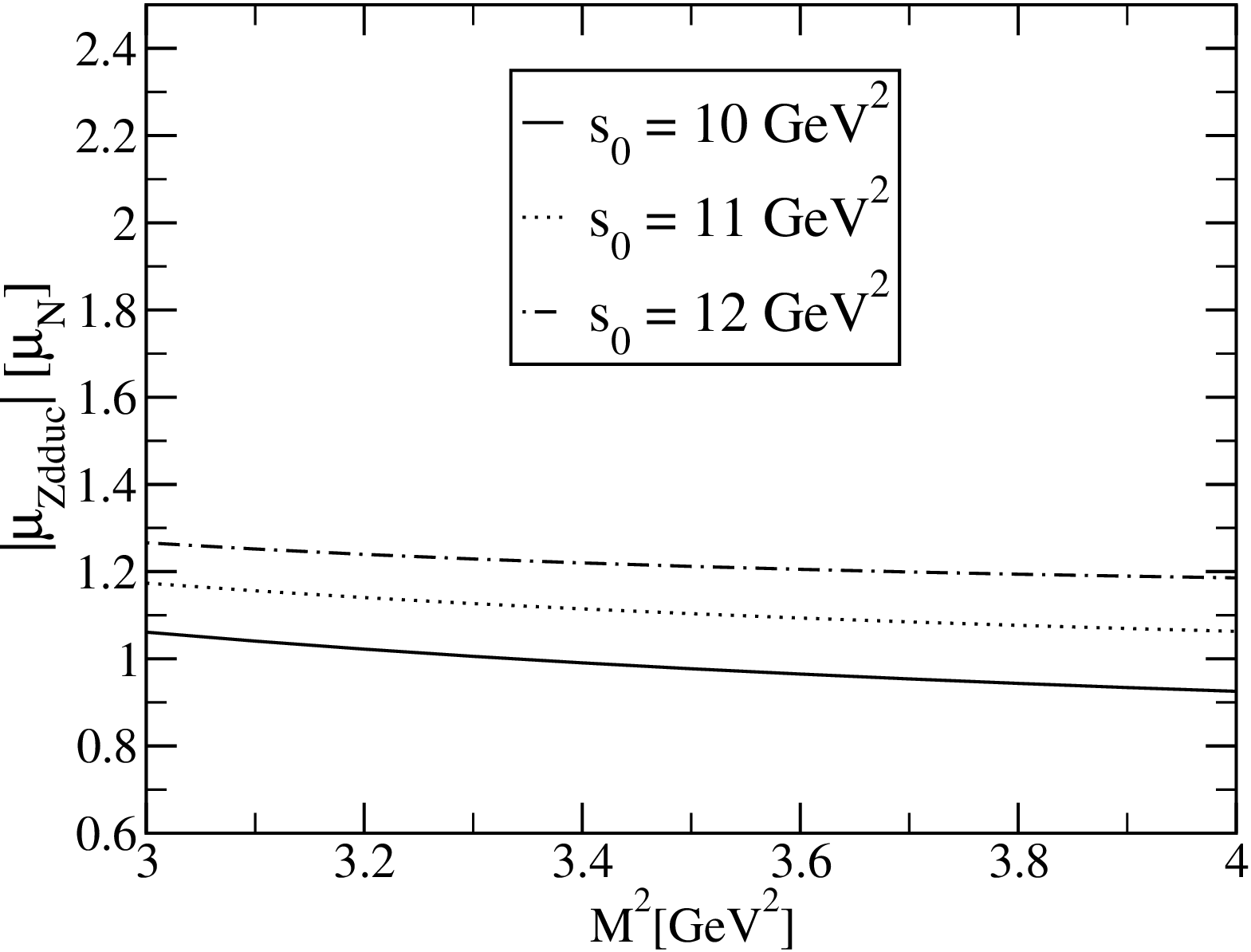}}~~~~~~~
 \subfloat[]{\label{fig:4QMMsq.eps}\includegraphics[width=0.35\textwidth]{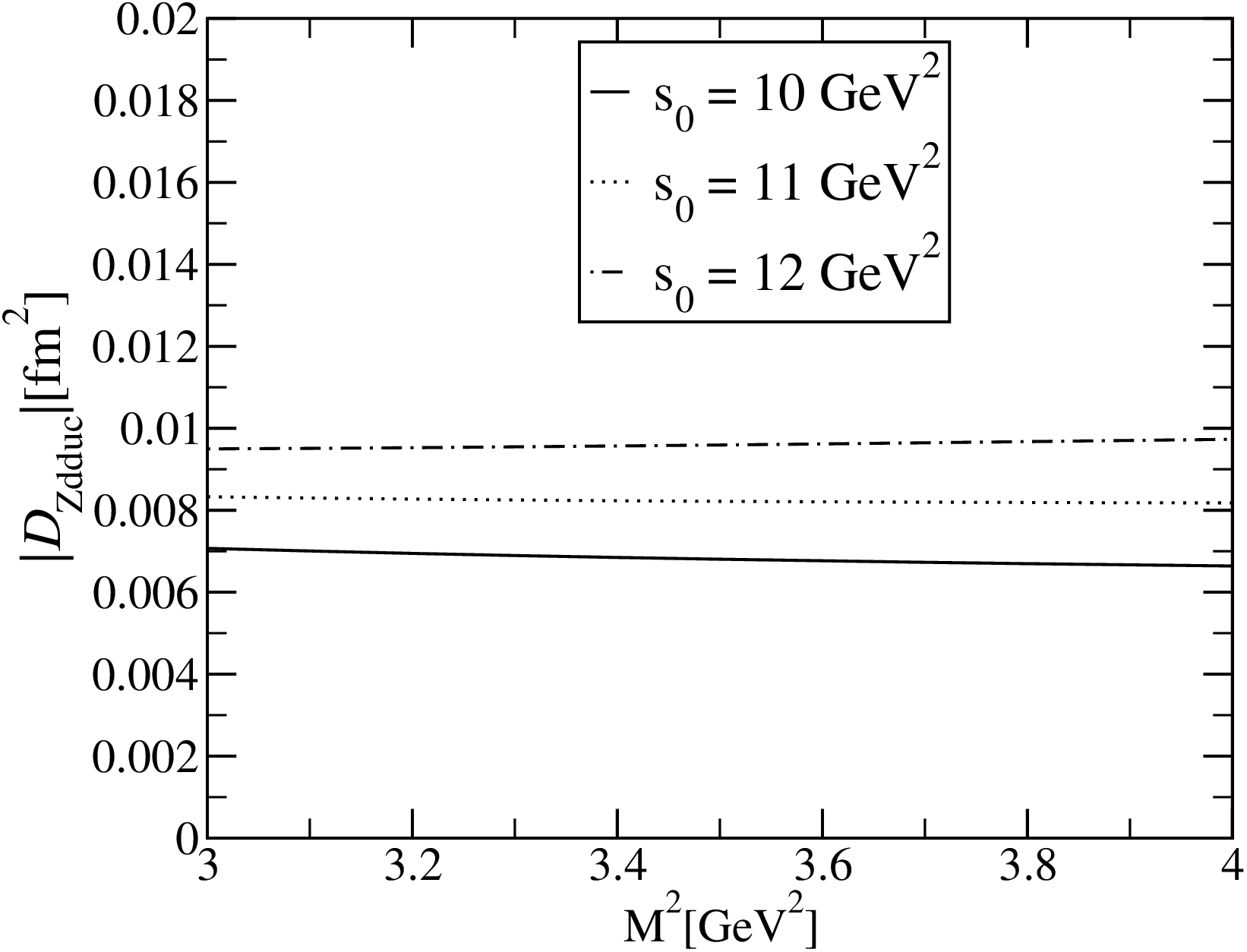}}\\
 \subfloat[]{\label{fig:5MMMsq.eps}\includegraphics[width=0.35\textwidth]{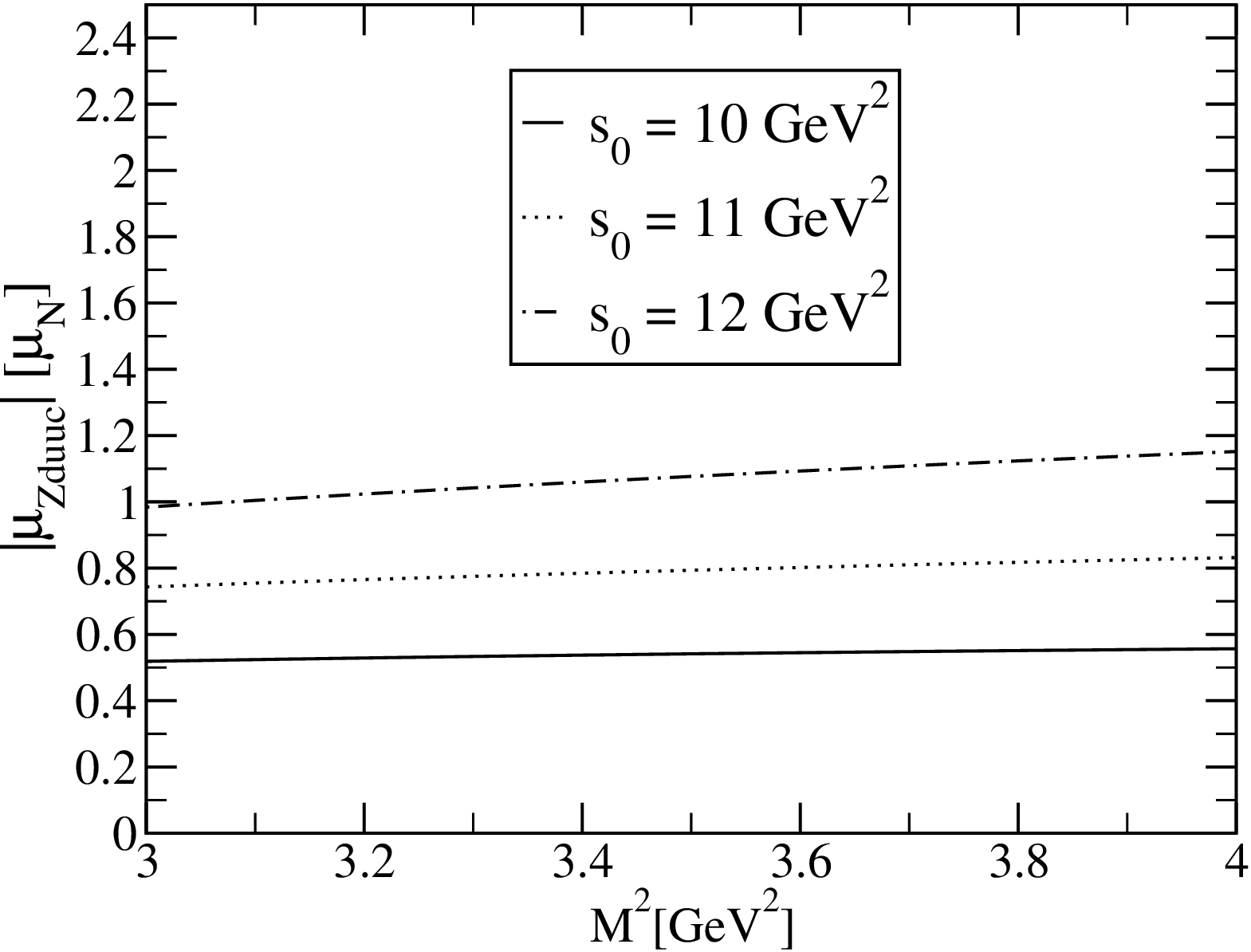}}~~~~~~~
 \subfloat[]{\label{fig:5QMMsq.eps}\includegraphics[width=0.35\textwidth]{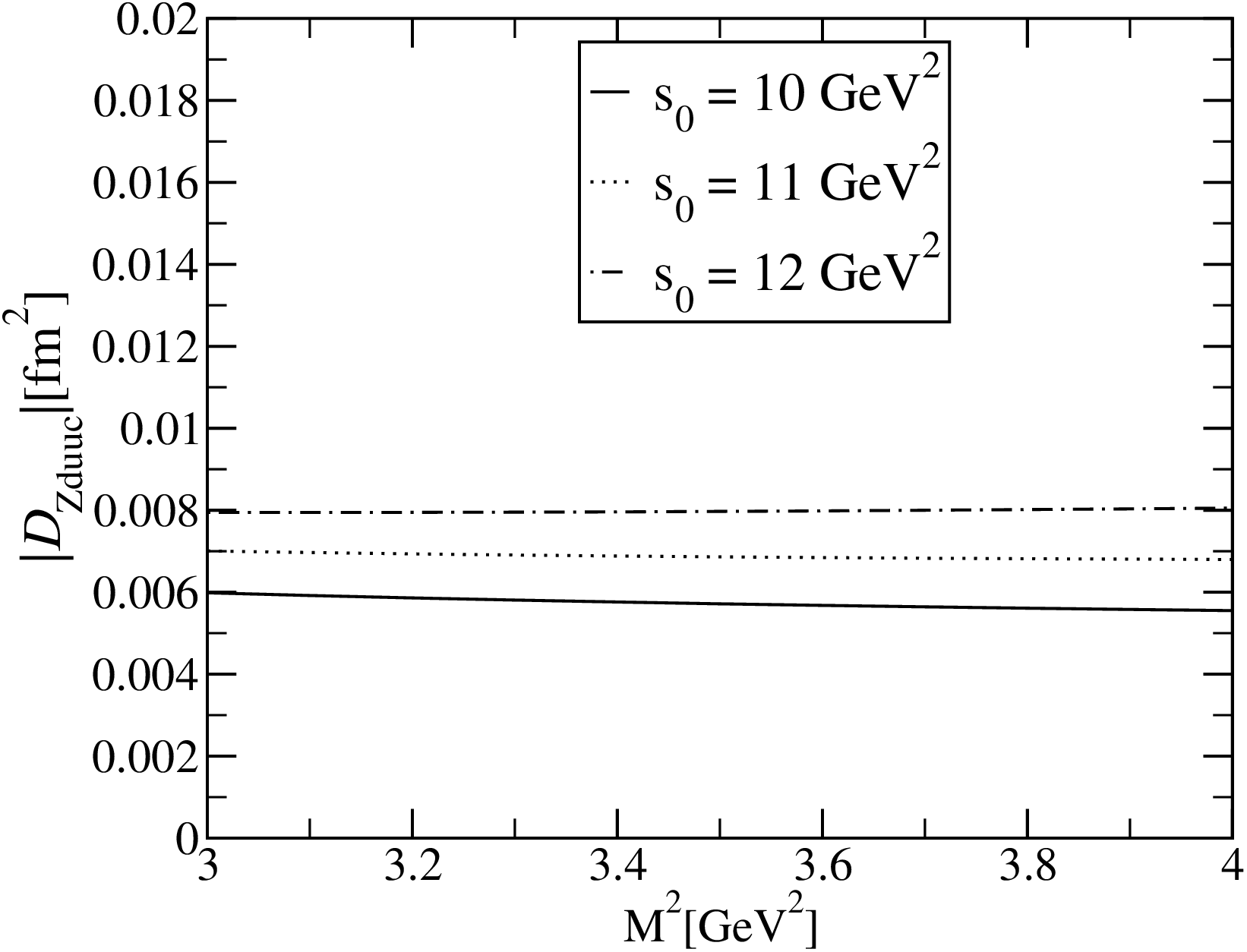}}\\
 \subfloat[]{\label{fig:6MMMsq.eps}\includegraphics[width=0.35\textwidth]{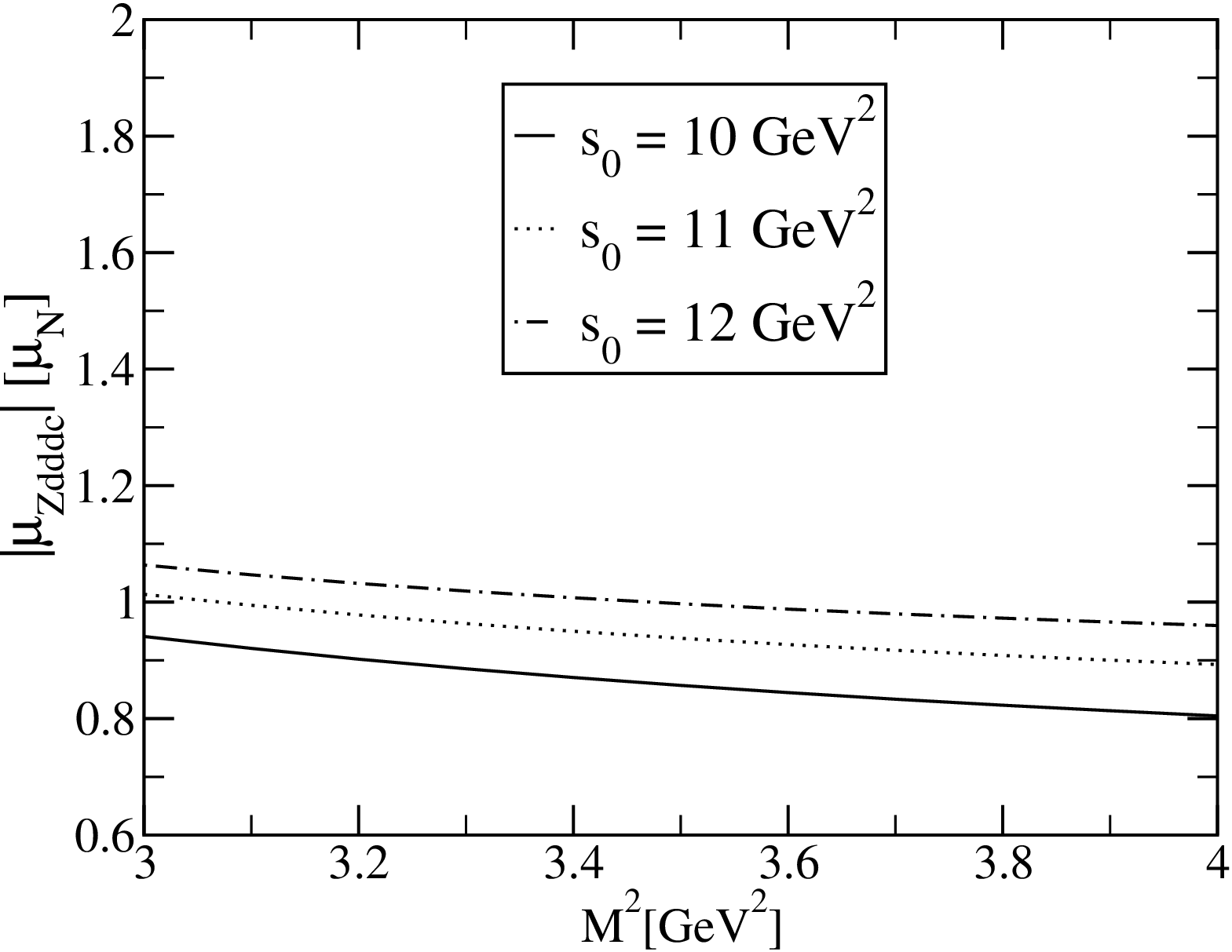}}~~~~~~~
 \subfloat[]{\label{fig:6QMMsq.eps}\includegraphics[width=0.35\textwidth]{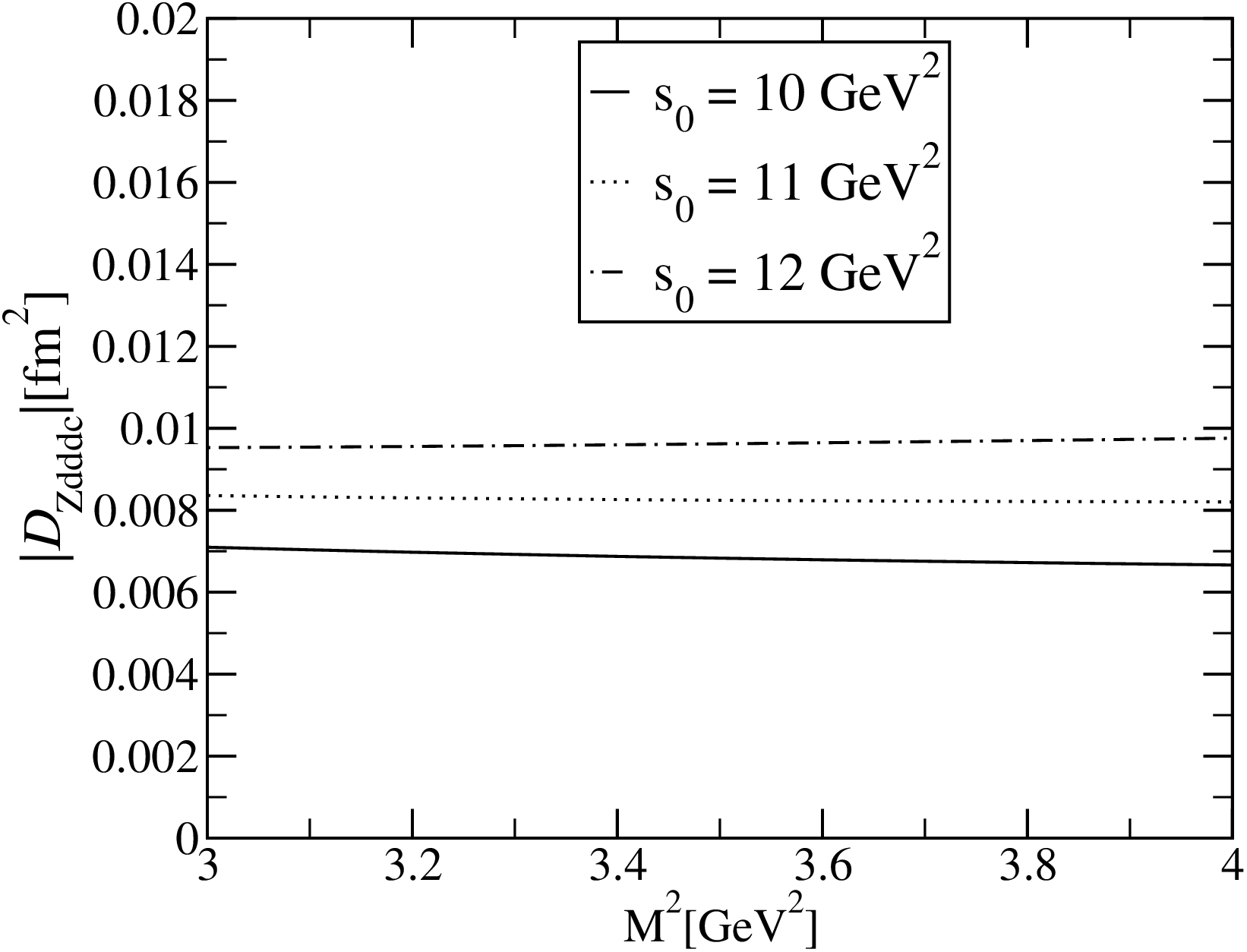}}\\
 \subfloat[]{\label{fig:7MMMsq.eps}\includegraphics[width=0.35\textwidth]{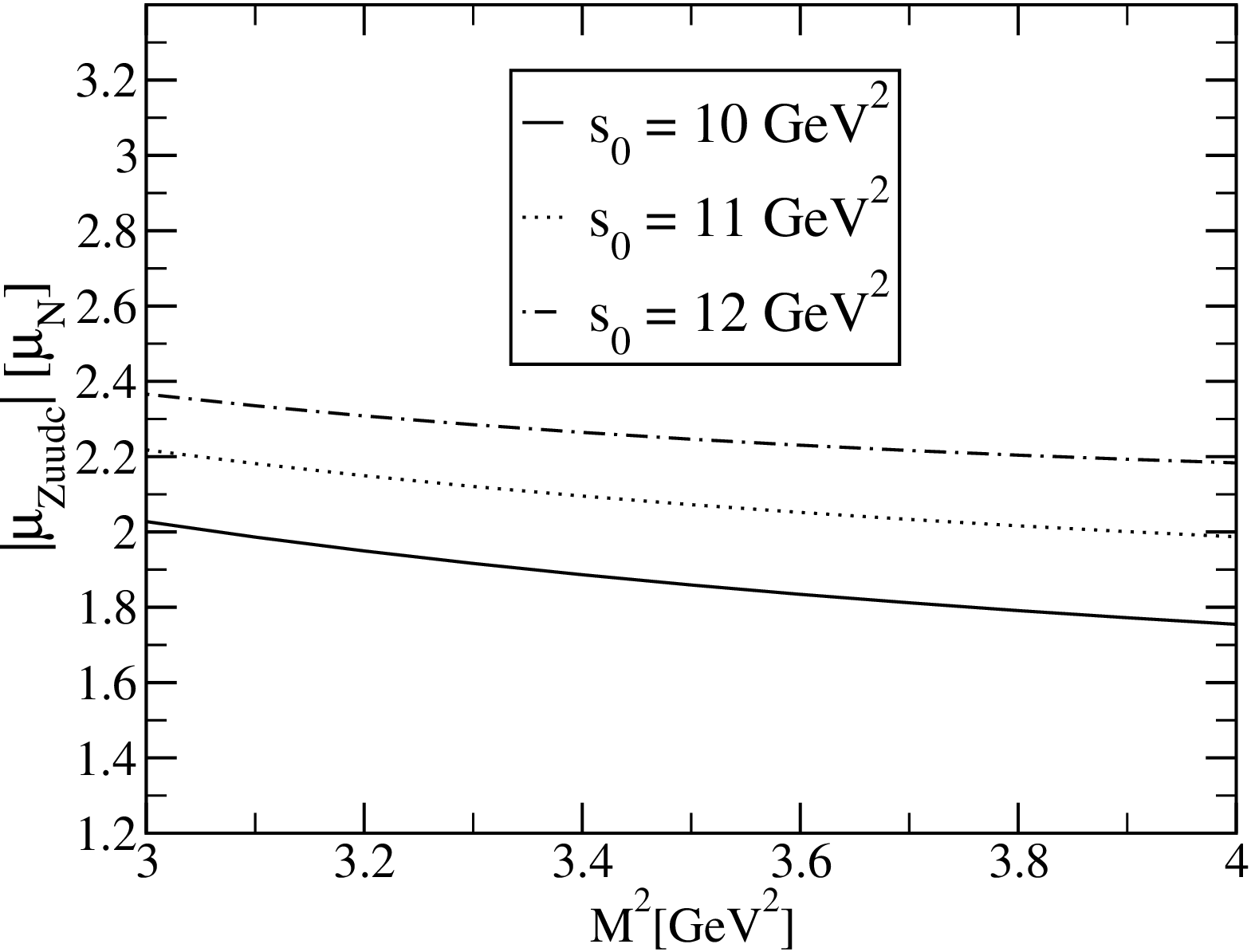}}~~~~~~~
 \subfloat[]{\label{fig:7QMMsq.eps}\includegraphics[width=0.35\textwidth]{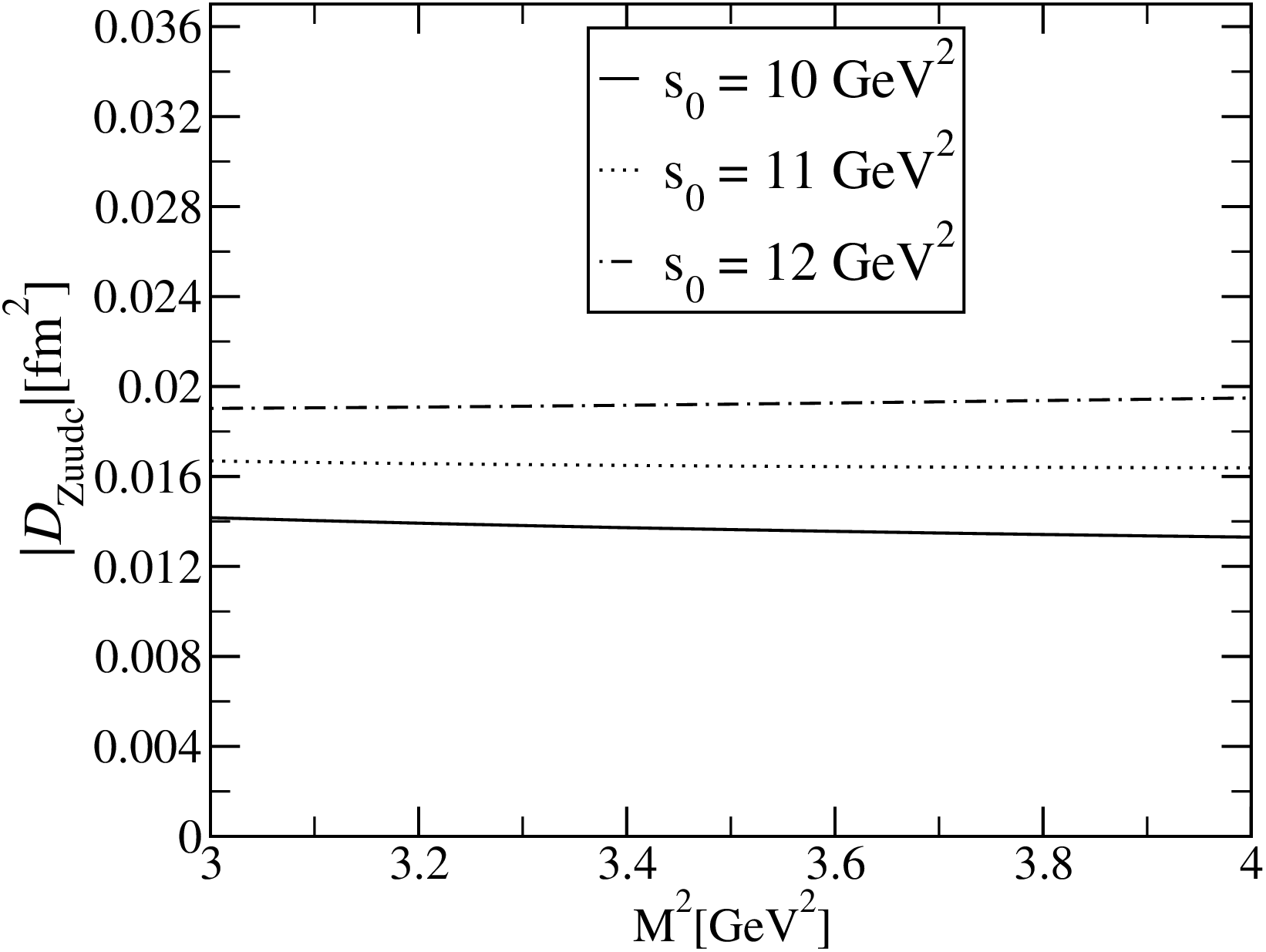}}\\
 \caption{ The dependence of the magnetic and quadrupole moments on the Borel parameter squared $M^{2}$
 at different fixed values of the continuum threshold: 
 (a) and (b) for the $Z_{dd \bar u \bar c}$ state, 
 (c) and (d) for the $Z_{du \bar u \bar c}$ state,
 (e) and (f) for the $Z_{dd \bar d \bar c}$ state and,
 (g) and (h) for the $Z_{uu \bar d \bar c}$ state.}
  \end{figure}
  
\section{Acknowledgement}

This work has been supported by the Scientific and
Technological Research Council of Turkey (T\"{U}B\.{I}TAK)
under the Grant No. 115F183.
\section*{Appendix A: Explicit forms of the functions \texorpdfstring{$\Pi_{i}^{QCD}$}{Lg}} 
In this appendix, we present the explicit expressions for the functions $\Pi_1^{QCD}$ and $\Pi_2^{QCD}$: 
\begin{align}
 \Pi_1^{QCD} &= \frac{1}{442368 \pi^2}\Bigg[32 P_1 \Bigg\{eq_3 mc P_3
 +144eq_1 \pi^2P_2 (m_0^2-m_c^2)
 +3eq_2\Bigg(m_{q_1} P_3 + 12 \Big(m_c^2 m_{q_1}+8\pi^2 m_c m_{q_1} P_1\nonumber\\
 &+4\pi^2m_0^2 P_2-8\pi^2 m_c^2 P_2\Big)\Bigg)\Bigg\}\mathbb{A}(u_0)
 +64 \chi P_1\Bigg\{8 eq_1 \pi^2 P_2\Bigg( P_3
 +6m_c\Big(-3m_c m_0^2+2m_c^3+16\pi^2P_1\Big)\Bigg)\nonumber\\
 &+eq_2\Bigg(P_3(3m_c^2m_{q_1}-8\pi^2 P_2)
 +48\pi^2 m_c \Big(3m_{q_1} m_0^2 P_1-3m_c m_0^2 P_2
 +2m_c^3P_2 +16 \pi^2 P_1 P_2\Big)
 \Bigg)\Bigg\}\varphi_\gamma(u_0)\nonumber\\
 &+32f_{3\gamma}\Bigg\{-eq_3m_c^2 P_3 
 +3eq_1 m_c\Big(m_c P_3 +96 \pi^2P_1(m_0^2-m_c^2)\Big)
 +3eq_2\Bigg(-m_c^2 P_3 - 48\pi^2 \Big(m_0^2 (2m_c P_1+m_{q_1} P_2)\nonumber\\
 & -m_c^2(2m_cP_1+3m_{q_1} P_2)\Big)\Bigg)\Bigg\}\psi^a(u_0)
 -1536\pi^2f_{3\gamma}\Bigg\{3eq_1m_cm_0^2 P_1
 +eq_2\Big(-3m_c^2m_{q_1} P_1+m_0^2(3m_cP_1\nonumber\\
 &+3m_{q_1} P_2)\Big)
 \Bigg\}\Bigg(\psi^\nu(u_0)+2I_6[\psi^\nu]\Bigg)
 +192P_1\Bigg\{24eq_1\pi^2 P_2 (m_0^2-2m_c^2)
 +eq_2\Bigg(m_{q_1} P_3 + 24\pi^2 \Big(4m_c m_{q_1} P_1\nonumber\\
 & +m_0^2 P_2 -2m_c^2 P_2\Big)\Bigg)\Bigg\}I_6[h_\gamma]
 +11eq_3m_c^2 f_{3\gamma}P_3 I_2[\mathcal{A}]
 +(eq_1-eq_2)m_c f_{3\gamma}\Big(23m_c P_3
 +576 \pi^2 P_1(m_0^2-2m_c^2)\Big)I_1[\mathcal{A}]\nonumber\\
 & +576eq_3 \pi^2m_c m_{q_1} P_1P_2\Big(2I_5[\mathcal{S}]
 +I_2[\mathcal{S}]\Big)
 -44eq_3m_c P_1 P_3\Big( I_4[\mathcal{T}_1]+ I_5[\mathcal{T}_1]\Big)\nonumber\\
 & -240eq_3 \pi^2 m_{q_1} f_{3\gamma} P_2 (m_0^2-3m_c^2)I_2[\mathcal{V}]
 +(eq_1+eq_2)m_cf_{3\gamma}\Big(-23m_c P_3 
 -576 \pi^2 P_1(m_0^2-2m_c^2)\Big)I_1[\mathcal{V}]\Bigg]I_7[0]\nonumber\\
 &+\frac{1}{221184 m_c^2 \pi^4}\Bigg[-16 m_c P_1\Bigg\{
 eq_3 P_3 +36m_c\Big(3eq_2 m_c^2 m_{q_1} 
 -8(eq_1+eq_2)\pi^2 P_2\Big)\Bigg\}\mathbb{A}(u_0)\nonumber\\
 &+192m_c^2 \chi P_1 \Big\{ -eq_2 m_{q_1} P_3
 +96 eq_2 \pi^2 m_c m_{q_1} P_1
 +24(eq_1+eq_2)\pi^2 P_1(m_0^2-2m_c^2)\Bigg\}\varphi_\gamma(u_0)\nonumber\\
 &+32m_cf_{3\gamma}\Bigg\{ (eq_3-3eq_1)m_c P_3
 -72 eq_1 \pi^2 P_1 (m_0^2-4m_c^2)
 +3eq_2\Bigg(m_c P_3 + 24 \pi^2 \Big(m_0^2 P_1 
 -4m_c(m_c P_1\nonumber\\
 &+m_{q_1} P_2)\Big)\Bigg)\Bigg\}\psi^a(u_0)
 +2304 \pi^2 m_c f_{3\gamma}\Bigg\{ (eq_1+eq_2) m_0^2 P_1
 -2eq_2 m_c m_{q_1} P_1\Bigg\}\Bigg(\psi^\nu(u_0)+2I_6[\psi^\nu]\Bigg)\nonumber\\
 &+96 P_1\Bigg\{ 96 (eq_1+eq_2)\pi^2 m_c^2 P_2
 +eq_2 m_{q_1} \Big(P_3 + 96 \pi^2m_c P_1\Big)\Bigg\}I_6[h_\gamma]
 +22eq_3m_c P_1 P_3 
 \Bigg( I_5[\mathcal{T}_1]+I_5[\mathcal{T}_2]\Bigg)\nonumber\\
 & -288eq_3 \pi^2 m_c m_{q_1} P_1P_2 I_2[\mathcal{S}]
 -11 eq_3m_c^2 f_{3\gamma} P_3 I_2[\mathcal{A}]
 +(eq_1-eq_2)m_c^2 f_{3\gamma}\Big(-23 P_3
 + 1152 \pi^2 m_c P_1 \Big) I_1[\mathcal{A}]\nonumber\\
 &-432 eq_3 \pi^2 m_c^2 m_{q_1} f_{3\gamma} P_2 I_2[\mathcal{V}]
 -(eq_1+eq_2)m_c^2 f_{3\gamma}\Big(-23 P_3
 + 1152 \pi^2 m_c P_1 \Big)I_1[\mathcal{V}]\Bigg]I_7[1]\nonumber\\
  &+\frac{1}{442368 m_c^2 \pi^4} \Bigg[ 3456 eq_2 m_c^2 m_{q_1} P_1 \mathbb{A}(u_0)
 +192 \chi P_1\Bigg\{96(eq_1+eq_2)\pi^2 m_c^2 P_2
 +eq_2 m_{q_1} \Big(P_3 + 96 \pi^2 m_c P_1\Big)\Bigg\} \varphi_\gamma(u_0)\nonumber\\
 & -32f_{3\gamma}\Bigg\{eq_3P_3+ 288eq_1 \pi^2 m_c P_1
 +3eq_2\Big(P_3 +48 \pi^2(2m_c P_1
 +ms P_2)\Big)\Bigg\}\psi^a(u_0)+11eq_3 f_{3\gamma} P_3 I_2[\mathcal{A}]\nonumber\\
 &-9216(eq_1+eq_2)\pi^2 P_1 P_2 I_6[h_\gamma]
 +2304 eq_2 \pi^2 m_{q_1} f_{3\gamma} P_2 I_3[\psi^a]
 +4608 eq_2 \pi^2 m_{q_1} f_{3\gamma} P_2 \Big( \psi^\nu(u_0)+2I_6[\psi^\nu]\Big)\nonumber\\
 & +(eq_1+eq_2)f_{3\gamma}\Big( 23 P_3
 +1152 \pi^2 m_c P_1 \Big) I_1[\mathcal{A}]
 +(eq_1+eq_2)f_{3\gamma}\Big(-23 P_3
 +1152 \pi^2 m_c P_1 \Big)I_1[\mathcal{V}]\nonumber\\ 
& +144eq_3\pi^2 m_{q_1} f_{3\gamma} P_2 I_2[\mathcal{V}]
\Bigg]I_7[2]\nonumber
\end{align}

\begin{align}
 &-\frac{eq_3 m_c m_{q_1} P_1 P_2 P_3}{20736 \pi^4}
 \Bigg[2 \varphi_\gamma(u_0)+I_3[\varphi_\gamma]\Bigg]I_7[-1]
 -\frac{P_1}{1152 m_c^2 \pi^4}\Bigg[
 16(eq_1+eq_2)\pi^2 \chi P_2 \varphi_\gamma(u_0)
 +3eq_2m_{q_1} \mathbb{A}(u_0)\Bigg]I_7[3]\nonumber\\
 &-\frac{m_c^4}{442368 \pi^4}\Bigg[
 64eq_3m_c \chi P_1 P_3 \varphi_\gamma(u_0)
 +64f_{3\gamma}\Bigg\{-eq3 P_3
 +3eq_1\Big(P_3+96 \pi^2 m_c P_1\Big)
 +3eq_2\Big(-P_3+48\pi^2\big(2m_c P_1\nonumber\\
 & -m_{q_1} P_2\big)\Big)\Bigg\}  \Big(2\psi^a(u_0)+I_3[\psi^a]\Big)
 -192 f_{3\gamma}\Bigg\{(eq_1+eq_2)\Big(P_3 +96 \pi^2 m_c P_1\Big)-48eq_2\pi^2m_{q_1} P_2\Bigg\}\Big(\psi^\nu(u_0)+2I_6[\psi^\nu]\Big)\nonumber\\
 & +18432 (eq_1+eq_2)\pi^2 P_1 P_2 I_6[h_\gamma]
 +32eq_3 m_c \chi P_3 P_1 I_3[h_\gamma]
 -11 eq_3 f_{3\gamma} P_3 I_2[\mathcal{A}]
 -288 eq_3 \pi^2 m_{q_1} f_{3\gamma}P_2 I_2[\mathcal{V}]\Bigg]I_8[-3,1]\nonumber\\
 &+\frac{m_c^2}{442368 \pi^2}\Bigg[
 -384 eq_2 m_{q_1} \chi P_1\Big( P_3
 +96 \pi^2 m_c P_1 \Big)\varphi_\gamma(u_0)
 -64 eq_3m_c \chi P_1P_3  I_3[\varphi_\gamma]-192(eq_1+eq_2)f_{3\gamma}\Big(P_3\nonumber\\
 & +96 \pi^2 m_c P_1 \Big)\Big( \psi^\nu(u_0)+2I_6[\psi^\nu]\Big)
 +32f_{3\gamma}\Bigg\{(3eq_1-3eq_2-eq_3)P_3+288 (eq_1-eq_2)\pi^2m_c P_1\Bigg\}I_3[\psi^a]\nonumber\\
 & -2(eq_1-eq_2)f_{3\gamma}\Big(23P_3
 +1152\pi^2m_c P_1\Big)I_1[\mathcal{A}]+2(eq_1+eq_2)f_{3\gamma}\Big(23P_3
 +1152\pi^2m_c P_1\Big)I_1[\mathcal{V}]\nonumber\\
& -33eq_3 f_{3\gamma}P_3 I_2[\mathcal{A}]
\Bigg]I_8[-2,1]\nonumber\\
 &+\frac{P_1}{110592 m_c^2 \pi^4}\Bigg[8eq_3 m_c P_3 A(u_0)
 +576(eq_1-eq_2)\pi^2 m_c f_{3\gamma}m_0^2\Big(2\psi^a(u_0)+I_3[\psi^a]\Big)
 +144eq_3 \pi^2 m_c m_{q_1} P_2 I_2[\mathcal{S}]\nonumber\\
 &-1152 (eq_1+eq_2)\pi^2 m_c f_{3\gamma}m_0^2 \Big(\psi^\nu(u_0)+2I_6[\psi^\nu]\Big)
 +48eq_2 m_{q_1} \Big(-P_3+96\pi^2 m_c P_1 \Big)I_6[h_\gamma]\nonumber\\
 &-11eq_3 m_c P_3 \Big(I_5[\mathcal{T}_1]-I_5[\mathcal{T}_2]\Big)
 \Bigg]I_8[0,0]\nonumber\\
 &-\frac{1}{221184 m_c^2 \pi^4}\Bigg[
 64 \chi P_1 \Bigg\{P_3(2eq_3m_c-3eq_2 m_{q_1})
 +288 \pi^2 m_c m_{q_1} P_1 \Bigg\}\varphi_\gamma(u_0)
 +32f_{3\gamma}\Bigg\{3eq_1\Big(96\pi^2 m_c P_1-P_3\Big)+eq_3 P_3 \nonumber\\
 & +3eq_2\Big(P_3  +48\pi^2 (2m_c P_1+m_{q_1} P_2)\Big)\Bigg\}
 \Big(\psi^a(u_0)-2I_3[\psi^a]\Big)
 -4608eq_2\pi^2 m_{q_1} f_{3\gamma}P_2\Big(\psi^\nu(u_0)+2I_6[\psi^\nu]\Big)\nonumber\\
 & -9216eq_2\pi^2 P_1 P_2 I_6[h_\gamma]\Bigg]I_8[0,1]\nonumber\\
 &+\frac{1}{1536 m_c^2 \pi^4}\Bigg[32eq_2 m_{q_1} \chi P_1 \varphi_\gamma(u_0)
+16(eq_1+eq_2)f_{3\gamma}\Big(\psi^\nu(u_0)+2I_6[\psi^\nu]\Big)
-8(eq_1-eq_2)f_{3\gamma}I_3[\psi^a]\nonumber\\
&+f{3\gamma}\Big\{2(eq_1-eq_2)I_1[\mathcal{A}]-eq_3I_2[\mathcal{V}]
-2(eq_1+eq_2)I_1[\mathcal{V}]\Big\}\Bigg]I_8[0,3]\nonumber\\
 &+\frac{m_c^4}{6144 \pi^4}\Bigg[64eq_2m_{q_1} \chi P_1\Big(m_c^2F[-4,3]+I_8[-3,3]\Big)\varphi_\gamma(u_0)
 -16(eq_1-eq_2)f_{3\gamma}\Big(m_c^4F[-5,3]+I_8[-3,3]\Big)\psi^a(u_0)\nonumber\\
  &+f_{3\gamma}\Bigg\{4(eq_1-eq_2)\Big(m_c^2F[-4,3]+I_8[-3,3]\Big)\Big(I_4[\mathcal{A}]-I_4[\mathcal{V}]\Big)
 +eq_3 \Big(m_c^4F[-5,3]-2m_c^2F[-4,3]+I_8[-3,3]\Big)I_2[\mathcal{V}]\nonumber\\
 &+16(eq_1+eq_2)f_{3\gamma}\Big(m_c^4F[-5,3]+2m_c^2F[-4,3]+I_8[-3,3]\Big)\Big(\psi^\nu(u_0)+2I_6[\psi^\nu]\Big)
-8(eq_1-eq2)\Big(m_c^4I_8[-5,3] \nonumber\\
&+2m_c^2I_8[-4,3]+I_8[-3,3]\Big)I_3[\psi^a]\Bigg\}\Bigg]\nonumber
\end{align}

\begin{align}
 &-\frac{eq_3 m_c \chi P_1P_3}{1382 \pi^4}
 \Bigg[2\varphi_\gamma(u_0)+I_3[\varphi_\gamma]\Bigg] I_8[-1,1]
 +\frac{eq_3 m_c P_1P_3}{27648 \pi^4} I_3[\mathbb{A}]I_8[1,0]
 +\frac{eq_3 m_c P_1}{1024 \pi^4}\Big(4I_5[\mathcal{S}]-I_2[\mathcal{S}]\Big)I_8[-2,2]\nonumber\\
 &-\frac{m_c^6 P_1}{1024 \pi^4}\Bigg[
 16 eq_2 m_{q_1} I_6[h_\gamma]+eq_3 m_c I_1[\mathcal{S}]\Bigg] I_8[-4,2]
 +\frac{P_1}{384 m_c^2 \pi^4}\Bigg[3eq_2 m_{q_1} \mathbb{A}(u_0)
 +16(eq_1+eq_2)\pi^2 \chi P_2 \varphi_\gamma(u_0)\Bigg]I_8[0,2] \nonumber\\
 &-\frac{m_c^6 m_{q_1} P_1}{512 \pi^4}\Bigg[4 eq_2 \Big(\mathbb{A}(u_0)
 -2 I_6[h_\gamma]\Big)
 - eq_3 \Big(I_2[\mathcal{S}]-2I_5[\mathcal{S}]\Big)\Bigg]I_8[-3,2].
\end{align}

and

\begin{align}
 \Pi_2^{QCD}&=-\frac{m_c^3 P_1}{55296 \pi^4}\Bigg[
 11eq_3P_3\Big(I_5[\mathcal{T}_1]+I_5[\mathcal{T}_2]\Big)
 -4608(eq_1+eq_2)\pi^2f_{3\gamma}m_0^2 I_6[\psi^\nu]\Bigg]I_7[-2]\nonumber\\
 &-\frac{f_{3\gamma}}{55296 \pi^4}\Bigg[
 11eq_3 P_3 I_5[\mathcal{A}]
 +9216eq_2\pi^2 m_{q_1} P_2 I_6[\psi^\nu]
 +(eq_1+eq_2)\Big(23 P_3
 +1152 m_c P_1\Big)I_4[\mathcal{A}]\Bigg]\Big(I_7[0]-m_c^2I_7[1]\Big)\nonumber\\
  &-\frac{f_{3\gamma}}{55296 m_c^4 \pi^4}\Bigg[11eq_3 P_3 I_5[\mathcal{A}]
 +(eq_1-eq_2)\Big(23 P_3+1152m_c P_1\Big)I_4[\mathcal{A}]
 +9216eq_2 \pi^2 m_{q_1} P_2 I_6[\psi^\nu]\Bigg]I_8[0,0]\nonumber\\
 &+\frac{f_{3\gamma}m_c^4}{128 \pi^4}\Bigg[
 (eq_1-eq_2)\Big(m_c^2I_8[-4,2]-I_8[-3,2]\Big)I_4[\mathcal{A}]
 +8(eq_1+eq_2)\Big(m_c^4I_8[-5,2]-2m_c^2I_8[-4,2]\nonumber\\
 &-I_8[-3,2]\Big)I_6[\psi^\nu]\Bigg]
 +\frac{11eq_3 P_1 P_3}{18432 \pi^4}
 \Bigg(I_5[\mathcal{T}_1]+I_5[\mathcal{T}_2]\Bigg)I_7[-1],
 \end{align}
where the values of $e_{q_1}$, $e_{q_2}$, $e_{q_3}$,  $m_{q_1}$,  $P_1$,  $P_2$  and  $P_3$ corresponding to different states are given in Table II.
 \begin{table}[htp]
	\addtolength{\tabcolsep}{10pt}
	\begin{center}
      \begin{tabular}{ccccccccc}
            \hline\hline
            ${Z_{\bar cq}}$          & $e_{q_1}$  & $e_{q_2}$ & $e_{q_3}$&$m_{q_1}$&  $P_1$&  $P_2$&  $P_3$& \\
            \hline\hline
	    $[sd][\overline{uc}]$    & $e_s$      &$e_d$      &  $e_u$   &$m_s$&$\langle \bar qq \rangle$
				     &$\langle \bar ss \rangle$&$\langle g_s^2 G^2 \rangle$& \\
	    $[sd][\overline{dc}]$    & $e_s$      &$e_d$      &  $e_d$   &$m_s$&$\langle \bar qq \rangle$
				     &$\langle \bar ss \rangle$&$\langle g_s^2 G^2 \rangle$&\\
	    $[su][\overline{uc}]$    & $e_s$      &$e_u$      &  $e_u$   &$m_s$&$\langle \bar qq \rangle$
				     &$\langle \bar ss \rangle$&$\langle g_s^2 G^2 \rangle$&\\
	    $[dd][\overline{uc}]$    & $e_d$      &$e_d$      &  $e_u$   &0&$\langle \bar qq \rangle$
				     &$\langle \bar qq \rangle$&$\langle g_s^2 G^2 \rangle$&\\
	    $[du][\overline{uc}]$    & $e_d$      &$e_u$      &  $e_u$   &0&$\langle \bar qq \rangle$
				     &$\langle \bar qq \rangle$&$\langle g_s^2 G^2 \rangle$&\\
	    $[dd][\overline{dc}]$    & $e_d$      &$e_d$      &  $e_d$   &0 &$\langle \bar qq \rangle$
				     &$\langle \bar qq \rangle$&$\langle g_s^2 G^2 \rangle$&\\
	    $[uu][\overline{dc}]$    & $e_u$      &$e_u$      &  $e_d$   &0 &$\langle \bar qq \rangle$
				     &$\langle \bar qq \rangle$&$\langle g_s^2 G^2 \rangle$&\\
	    \hline\hline
\end{tabular}
\end{center}
\caption{The values of $e_{q_1}$, $e_{q_2}$, $e_{q_3}$,  $m_{q_1}$,  $P_1$,  $P_2$  and  $P_3$
related to the expressions of the magnetic and quadrupole moments in Eqs.(17) and (18).}
	\label{table2}
\end{table}

The functions~$I_1[\mathcal{A}]$,~$I_2[\mathcal{A}]$,~$I_3[\mathcal{A}]$,~$I_4[\mathcal{A}]$,~$I_5[\mathcal{A}]$,~$I_6[\mathcal{A}]$,~$I_7[n]$ and~$I_8[n,m]$  are
defined as:
\begin{align}
 I_1[\mathcal{A}]&=\int D_{\alpha_i} \int_0^1 dv~ \mathcal{A}(\alpha_{\bar q},\alpha_q,\alpha_g)
 \delta'(\alpha_ q +\bar v \alpha_g-u_0),\nonumber\\
  I_2[\mathcal{A}]&=\int D_{\alpha_i} \int_0^1 dv~ \mathcal{A}(\alpha_{\bar q},\alpha_q,\alpha_g)
 \delta'(\alpha_{\bar q}+ v \alpha_g-u_0),\nonumber\\
  I_3[\mathcal{A}]&=\int_0^1 du~ A(u)\delta'(u-u_0),\nonumber\\
  I_4[\mathcal{A}]&=\int D_{\alpha_i} \int_0^1 dv~ \mathcal{A}(\alpha_{\bar q},\alpha_q,\alpha_g)
 \delta(\alpha_ q +\bar v \alpha_g-u_0),\nonumber\\
   I_5[\mathcal{A}]&=\int D_{\alpha_i} \int_0^1 dv~ \mathcal{A}(\alpha_{\bar q},\alpha_q,\alpha_g)
 \delta(\alpha_{\bar q}+ v \alpha_g-u_0),\nonumber
 \end{align}
 \begin{align}
 I_6[\mathcal{A}]&=\int_0^1 du~ A(u),\nonumber\\
 I_7[n]&=  \int_{m_c^2}^{s_0} ds~e^{-s/M^2}~s^n~,\nonumber\\
 I_8[n,m]&= \int_{m_c^2}^{s_0} ds \int_{m_c^2}^s dl~ e^{-s/M^2}~\frac{(s-l)^m}{l^n}\nonumber.
 \end{align}

\bibliography{refs}

\end{document}